\documentclass[twocolumn,english,reprint, longbibliography, superscriptaddress, breaklinks=true, showkeys, showpacs=false, nofootinbib]{revtex4}
\usepackage[T1]{fontenc}
\usepackage[utf8]{inputenc} 
\usepackage{color}
\usepackage{babel}
\usepackage{amsmath}
\usepackage{amssymb}
\usepackage{subfigure}
\usepackage{graphicx}
\usepackage{physics}
\usepackage{appendix}
\newtheorem{theorem}{Theorem}

\newtheorem{lemma}{Lemma}

\usepackage[unicode=true,pdfusetitle,
 bookmarks=true,bookmarksnumbered=false,bookmarksopen=false,breaklinks=true,pdfborder={0 0 0},backref=false,colorlinks=true]
 {hyperref} 
\makeatletter
\@ifundefined{textcolor}{}
{%
 \definecolor{BLACK}{gray}{0}
 \definecolor{WHITE}{gray}{1}
 \definecolor{RED}{rgb}{1,0,0}
 \definecolor{GREEN}{rgb}{0,1,0}
 \definecolor{BLUE}{rgb}{0,0,1}
 \definecolor{CYAN}{cmyk}{1,0,0,0}
 \definecolor{MAGENTA}{cmyk}{0,1,0,0}
 \definecolor{YELLOW}{cmyk}{0,0,1,0}
}
\pdfoutput=1
\hypersetup{colorlinks=true,citecolor=blue,linkcolor=cyan,urlcolor=blue,filecolor= green, breaklinks=true}
\usepackage{url}
\usepackage{breakurl}
\makeatother

\begin{document}

\title{The quantum cost function concentration dependency \\ on the parametrization expressivity}

\author{Lucas Friedrich}
\email[Electronic address: ]{lucas.friedrich@acad.ufsm.br}
\affiliation{Physics Departament, Center for Natural and Exact Sciences, Federal University of Santa Maria, Roraima Avenue 1000, 97105-900, Santa Maria, RS, Brazil}

\author{Jonas Maziero}
\email[Electronic address: ]{jonas.maziero@ufsm.br}
\affiliation{Physics Departament, Center for Natural and Exact Sciences, Federal University of Santa Maria, Roraima Avenue 1000, 97105-900, Santa Maria, RS, Brazil}

\selectlanguage{english}%

\begin{abstract}


Although we are currently in the era of noisy intermediate scale quantum devices, several studies are being conducted with the aim of bringing machine learning to the quantum domain. Currently, quantum variational circuits are one of the main strategies used to build such models. However, despite its widespread use, we still do not know what are the minimum resources needed to create a quantum machine learning model. In this article, we analyze how the expressiveness of the parametrization affects the cost function. We analytically show that the more expressive the parametrization is, the more the cost function will tend to concentrate around a value that depends both on the chosen observable and on the number of qubits used. For this, we initially obtain a relationship between the expressiveness of the parametrization and the mean value of the cost function. Afterwards, we relate the expressivity of the parametrization with the variance of the cost function. Finally, we show some numerical simulation results that confirm our theoretical-analytical predictions.

\end{abstract}


\maketitle

\section{Introduction}

In recent years, there has been a great increase in interest in quantum computing due to its possible applications in solving problems such as simulation of quantum systems \cite{quantum_simulation}, development of new drugs \cite{drug_discovery}, and resolution of systems of equations linear \cite{linear_system}. Quantum machine learning, which is an interdisciplinary area of study between machine learning and quantum computing, is also another possible application that should benefit from the computational power of these devices. In this sense, several models have already been proposed, such as Quantum Multilayer Perceptron \cite{quantum_model_multilayer_perception}, Quantum Convolutional Neural Networks \cite{qunatum_Convolutional}, Quantum Kernel Method \cite{kernel_methods}, and Quantum-Classical Hybrid Neural Networks \cite{hybrid_1,hybrid_2,hybrid_3,hybrid_4}. However, in the era of noisy intermediate scale quantum devices (NISQ), variational quantum algorithms (VQAs) \cite{VQA} are the main strategy used to build such models.

Variational quantum algorithms are models that use a classical optimizer to minimize a cost function by optimizing the parameters of a  parametrization $U$. Several optimization strategies have already been proposed \cite{Friedrich_es,Anand,Zhou_leo,fosel_thomas}, although this is an open area of study. In fact, despite the widespread use of VQAs, our understanding of VQAs is limited and some problems still need to be solved, such as the disappearance of the gradient \cite{McClean,BR_cost_Dependent,BR_Entanglement_devised_barren_plateau_mitigation,BR_Entanglement_induced_barren_plateaus,BR_expressibility,BR_noise,BR_gradientFree}, methods to mitigate the Barren Plateaus issue \cite{FRIEDRICH,BR_initialization_strategy,BR_Large_gradients_via_correlation,BR_LSTM,BR_layer_by_layer}, how to build a parameterization $U$ \cite{Quantum_architecture_Kuo,Friedrich_Restricting}, and how correct errors \cite{Quantum_error_mitigation}.

In this sense, in this article we aim to analyze how the expressivity of the parametrization $U$ affects the cost function. We will show that the more expressive the $U$ parametrization is, the more the average value of the cost function will concentrate around a fixed value. In addition, we will also show that the probability of the cost function deviating from its average will also depend on the quantum circuit expressivity.

The remainder of this article is organized as follows. In Section \ref{sec:vqa}, we make a short introduction about VQAs. In  Section \ref{sec:expressibility}, we comment on how expressiveness can be quantified and what is its meaning. In the following section, Section \ref{sec:main}, we present our main results. There we will give two theorems. In Theorem \ref{tr:concentracao}, we obtain a relationship between the concentration of the cost function and the expressiveness of the parametrization. In Theorem \ref{tr:variancia}, we obtain the probability for the cost function to deviate from its average value, restricting it via a function of the quantum circuit expressivity. Then, in Section \ref{sec:result}, we present some numerical simulation results to confirm our theoretical analytical predictions. Finally, Section \ref{sec:conclusion} presents our conclusions.

\section{Variational quantum algorithms}\label{sec:vqa}

\begin{figure}
    \centering
    \includegraphics[scale=0.3]{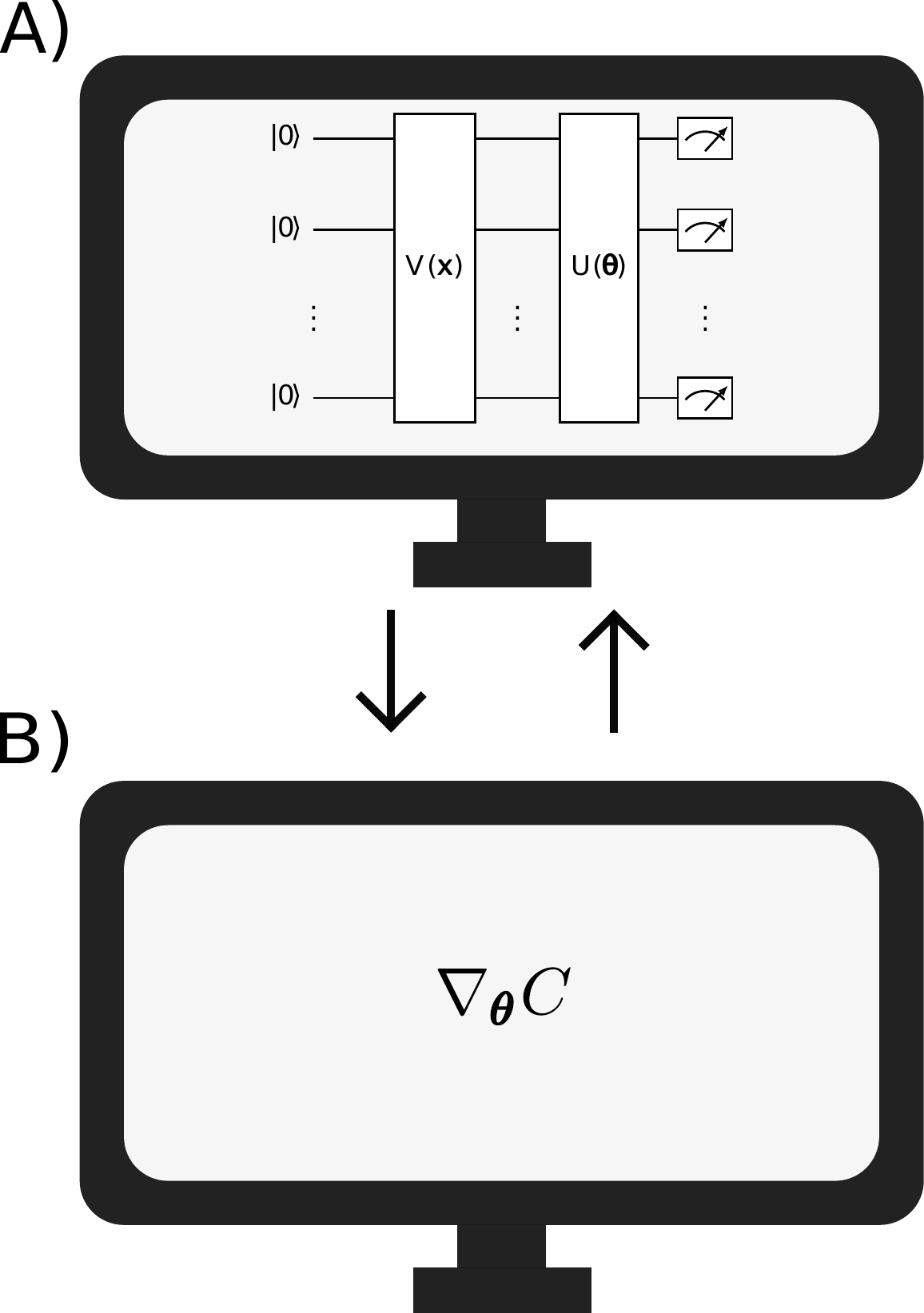}
    \caption{Illustration of how a quantum variational algorithm works. These models have two parts. A) Quantum circuit that runs on the quantum computer. B) Classical computer that optimizes the parameters, in general, using the gradient and the cost function. }
    \label{fig:model_vqa}
\end{figure}

Variational quantum algorithms are models where a classical optimizer is used to minimize a cost function, which is usually written as the average value of an observable $O$:
\begin{equation}
    C = Tr[OU(\pmb{\theta})|\psi \rangle \langle \psi|U(\pmb{\theta})^{\dagger}],\label{eq:cost_function}
\end{equation}
where $|\psi \rangle := V(\pmb{x})|0 \rangle$.
To do so, the optimizer updates the parameters $\pmb{\theta}$ of the parameterization $U$.
In Fig. \ref{fig:model_vqa}, one can see a schematic representation of how a VQA works. In the first part, Fig. \ref{fig:model_vqa} A, a quantum circuit runs on a quantum computer. In general, this circuit is divided into three parts. In the first part we have a $V$ parametrization that is used to encode data in a quantum state. In quantum machine learning, this parametrization is used to bring our data, such as data from the MNIST \cite{mnist_dataset} dataset, into a quantum state. Next, we have the parametrization $U$ that will depend on the parameters $\pmb{\theta}$ that we must optimize. Finally, we have the measures that are used to calculate the cost function. 
In the second part, Fig. \ref{fig:model_vqa} B, we have a classical computer that performs the task of optimizing the parametrization parameters. In general, for this task the gradient of the cost function is used.

In this article, the parametrization will be given by
\begin{equation}
    U = \prod_{l=1}^{L}U_{l} = \prod_{l=1}^{L}U'_{l}W_{l},\label{eq:parametrization}
\end{equation}
where $L$ is the number of layers, $U'_{l}$ is a layer that depends on the parameters $\pmb{\theta}$ and $W_{l}$ is a layer that does not depend on the parameters $ \pmb{\theta}$. The construction of parametrizations is still an open area of study and, due to the complexity involved in its construction, some works have proposed using the automation of this process \cite{Quantum_architecture_Kuo,Zhang_Shi_Xin}. Furthermore, for problems such as quantum machine learning, where a $V$ parameterization is used to encode data in a quantum state, the choice of $V$ is also extremely important \cite{Schuld_data}, and several possible encoding forms have been proposed \cite{LaRose_Ryan}.

\section{Expressivity}
\label{sec:expressibility}

Following Ref. \cite{Sim_Expressibility}, here we define expressivity as the ability of a quantum circuit to generate (pure) states that are well representative of the Hilbert space. In the case of a qubit, this comes down to the quantum circuit's ability to explore the Bloch sphere. To quantify the expressiveness of a quantum circuit, we can compare the uniform distribution of units obtained from the set $\mathbb{U}$ with the maximally expressive (Haar) uniform distribution  of units of $\mathcal{U}(d)$. For this, we use the following super-operator \cite{BR_expressibility}
\begin{equation}
    A_{\mathbb{U}}^{t}(.)  := \int_{\mathcal{U}(d)}d\mu (V)V^{\otimes t}(.)(V^{\dagger})^{\otimes t} 
     - \int_{\mathbb{U}}dUU^{\otimes t}(.)(U^{\dagger})^{\otimes t},
     \label{eq:expr_definition}
\end{equation}
where $d\mu(V)$ is a volume element of the Haar measure and $dU$ is a volume element corresponding to the uniform distribution over $\mathbb{U}$. 
The uniform distribution over $\mathbb{U}$ is obtained by fixing the parameterization $U$, where for each vector of parameters $\pmb{\theta}$ we obtain a unit $U(\pmb{\theta})$. Thus, given the set of parameters $\{ \pmb{\theta}^{1},\pmb{\theta}^{2},..., \pmb{\theta}^{m} \}$, we obtain the corresponding set of unitary operators:
\begin{equation}
    \mathbb{U} = \{ U^{1},U^{2}, ..., U^{m} \}.\label{eq:conjunto_U}
\end{equation}

\section{Main Theorems}
\label{sec:main}

In this section, we present our main results. First, we  obtain a relationship between the average value of the cost function, Eq. \eqref{eq:cost_function}, with the expressivity of the parametrization $U$, Eq. \eqref{eq:parametrization}. Afterwards, we will obtain a relationship between the variance of the cost function and the expressiveness of the parametrization. To do so, we start by writing the average of the cost function as
\begin{equation}
     E_{\mathbb{U}}[C] = \int_{\mathbb{U}} dU Tr[OU\rho U^{\dagger}].\label{eq:media_cost_function}
\end{equation}
Therefore, using Eq. \eqref{eq:expr_definition} in Eq. \eqref{eq:media_cost_function}, we obtain the following relationship between the mean of the cost function and the expressivity of the parametrization, Theorem \ref{tr:concentracao}.

\begin{theorem}\label{tr:concentracao}
    (Concentration of the cost function). Let the cost function be defined as in Eq. \eqref{eq:cost_function}, with observable $O$, parameterization $U$, Eq. \eqref{eq:parametrization}, and encoding quantum state $\rho := |\psi \rangle \langle \psi|$. Then it follows thus that
    \begin{equation}
        \bigg| E_{\mathbb{U}}[C] - \frac{Tr[O]}{d} \bigg| \leqslant \| O \|_{2} \| A(\rho) \|_{2}.\label{eq:eq_teo_1}
    \end{equation}
\end{theorem}

The proof of this theorem is presented in Appendix \ref{ap:Appendix A}.

Above we used the matrix $2$-norm $||A||_2^2 = Tr(A^\dagger A)$. For any operator $X$, from Eq. \eqref{eq:expr_definition}, we have that the smaller $\|A_{\mathbb{U}}^{t}(X)\|_{2}$, with $A^{t}_{\mathbb{U}}\equiv A$, the more expressive will be the parametrization. Therefore, Theorem \ref{tr:concentracao}  implies that the greater the expressiveness of the parameterization $U$, the more the cost function average will tend to have the value $Tr[O]/d$.

Despite Theorem \ref{tr:concentracao}  implying a tendency of the mean value of the cost function to go a fixed value, when executing the VQA, the cost function may deviate from its mean. To calculate this deviation we use the Chebyshev inequality,
\begin{equation}
    P(|C-E_{\mathbb{U}}[C]| \geqslant \delta  ) \leqslant \frac{  Var_{ \mathbb{U} }[C]  }{  \delta^{2} }, \label{eq:prob} 
\end{equation}
which informs the probability for the cost function to deviate from its mean value.

Next, we present the Theorem \ref{tr:variancia}, relating the modulus of the cost function variance with the expressiveness of the parametrization.

\begin{theorem}\label{tr:variancia}
    Let us consider the cost function defined in Eq. \eqref{eq:cost_function} and the parameterization $U$ defined in Eq. \eqref{eq:parametrization}. The variance of the cost function can be constrained as follows:
    \begin{equation}
        \begin{split}
        |Var_{\mathbb{U}}[C]| \leqslant &|\beta| +  \| O^{\otimes 2} \|_{2} \| A^{\otimes 2}(\rho^{\otimes 2}) \|_{2} + |\alpha|  \| O \|_{2} \| A(\rho) \|_{2}  \\ & +  \| O \|_{2}^{2} \| A(\rho) \|_{2}^{2},
        \label{eq:var_teorico}
        \end{split}
    \end{equation}
     with
     \begin{equation}
         \beta := \frac{ Tr[O]^{2} + Tr[O^{2}] }{d^{2}-1}\bigg( 1 - \frac{1}{d} \bigg)-\frac{Tr[O]^{2}}{d^{2}}
     \end{equation}
     and
     \begin{equation}
         \alpha := \frac{2Tr[O]}{d}.
     \end{equation}
     Here $d=2^{n}$, where $n$ is the number of qubits.
\end{theorem}

The proof of this theorem is presented in Appendix \ref{ap:Appendix B}.

As the variance is a positive real number, we can use Theorem \ref{tr:variancia} to analyze the probability that the cost function deviates from its mean, Eq. \eqref{eq:prob}. Therefore, from Theorem \ref{tr:variancia}, we see that by defining the observable $O$ and the size of the system, that is, the number of qubits used, the probability of the cost function deviating from its mean decreases as the expressivity increases. Furthermore, it also follows, from Theorem \ref{tr:concentracao}, that for maximally expressive parameterizations, i.e., for $\|A_{\mathbb{U}}^{t}(X)\|_{2} = 0 $, the cost function will be stuck to the fixed value $\frac{Tr[O]}{d} $.

\section{Simulation Results}
\label{sec:result}

In this section we will present some numerical simulation results. For this, we use twelve different parametrizations, which we call, respectively, Model 1, Model 2, ..., Model 12. See Appendix \ref{ap:Appendix C} for the corresponding quantum circuits. As we saw in Eq. \eqref{eq:parametrization}, the parametrization is obtained from the product of $L$ layers $U_{l}$, where each layer $U_{l}$ can be distinct from one another, that is, the gates and sequences we use in one layer may be different from another. However, in general, they are the same. For the results shown here, the $U_{l}$ layers are the same, the only difference being the $\pmb{\theta}$ parameters used in each layer.

For these results we define each $U'_{l}$ as
\begin{equation}
    U'_{l} :=  \bigotimes_{i=1}^{n}R_{Y}(\theta_{i,l}),\label{eq:U_model_fixo}
\end{equation}
where the index $l$ indicates the layer and the index $i$ the qubit. Also, we use $R_{Y}(\theta_{i,l}) = e^{-i \theta_{j,i} Y/2} $ in all models. In the parametrizations of Model 3, Model 4, Model 6, Model 9, Model 10, and Model 12, Figs. \ref{fig:figura_model_3}, \ref{fig:figura_model_4}, \ref{fig:figura_model_6}, \ref{fig:figura_model_9}, \ref{fig:figura_model_10}, and \ref{fig:figura_model_12}, respectively, before for each $U'_{l}$, we apply the Hadamard gate to all the qubits. Furthermore, in Model 2, Model 3, Model 8, and Model 9, Figs. \ref{fig:figura_model_2}, \ref{fig:figura_model_3}, \ref{fig:figura_model_8}, and \ref{fig:figura_model_9}, respectively, we use the controlled port $R_{Y}$, or $CRY$. Finally, for the results obtained here, we used the PennyLane \cite{Pennylane_bibli} library. Furthermore, the codes used to obtain these results are available for access in \url{https://github.com/lucasfriedrich97/quantum-expressibility-vs-cost-function}.

\begin{figure*}
    \centering
    \includegraphics[scale=0.40]{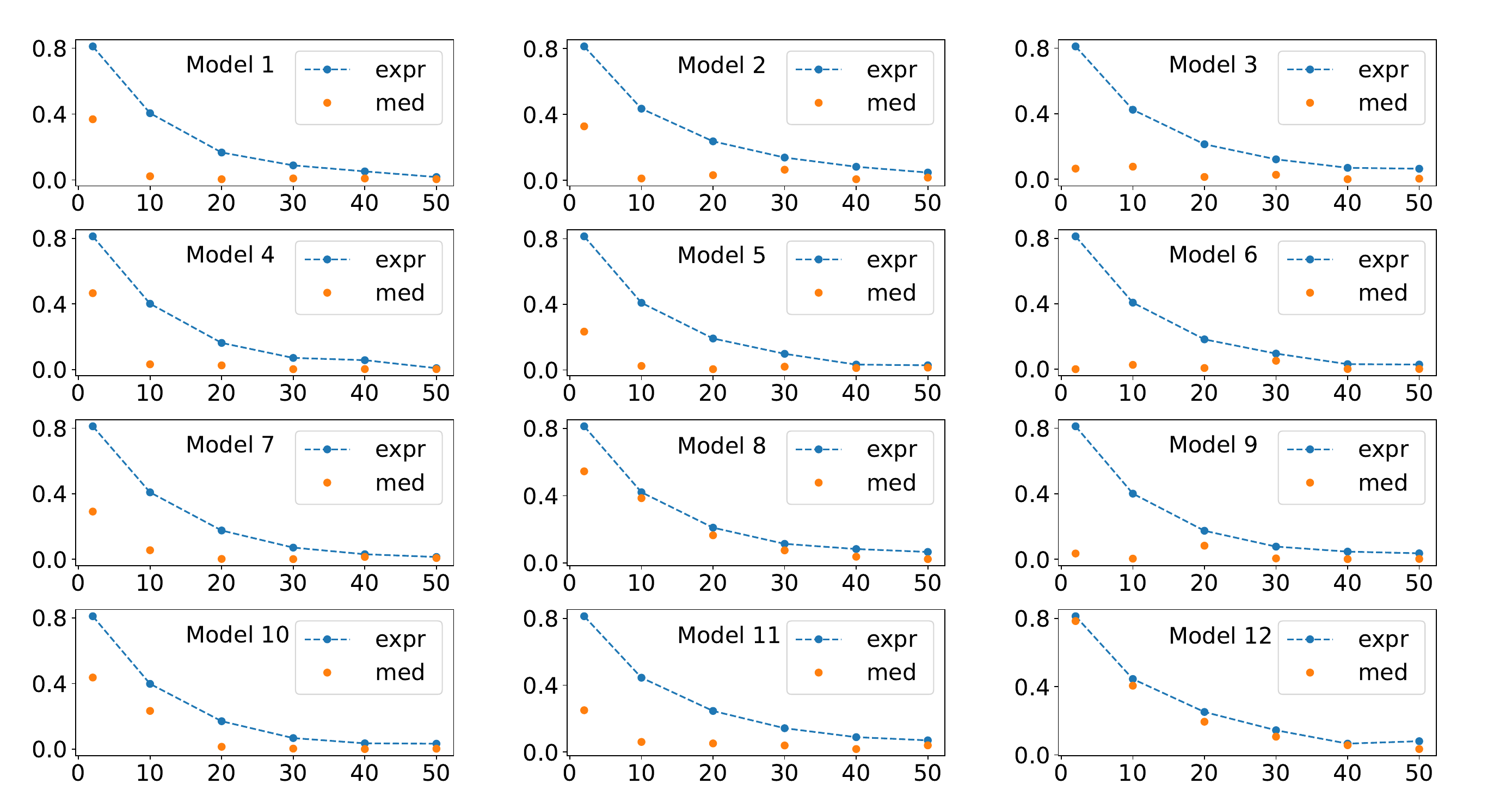}
    \caption{
    Behavour of the right hand side of Eq. (\ref{eq:eq_teo_1}), the quantum expressivity (expr), and of the average cost function term (med), the left hand side of Eq. (\ref{eq:eq_teo_1}),   
    as the number of layers $L$ is increased. Four qubits were used for obtaining all these plots.}
    \label{fig:teo_1_4_qubit}
\end{figure*}
\begin{figure*}
    \centering
    \includegraphics[scale=0.40]{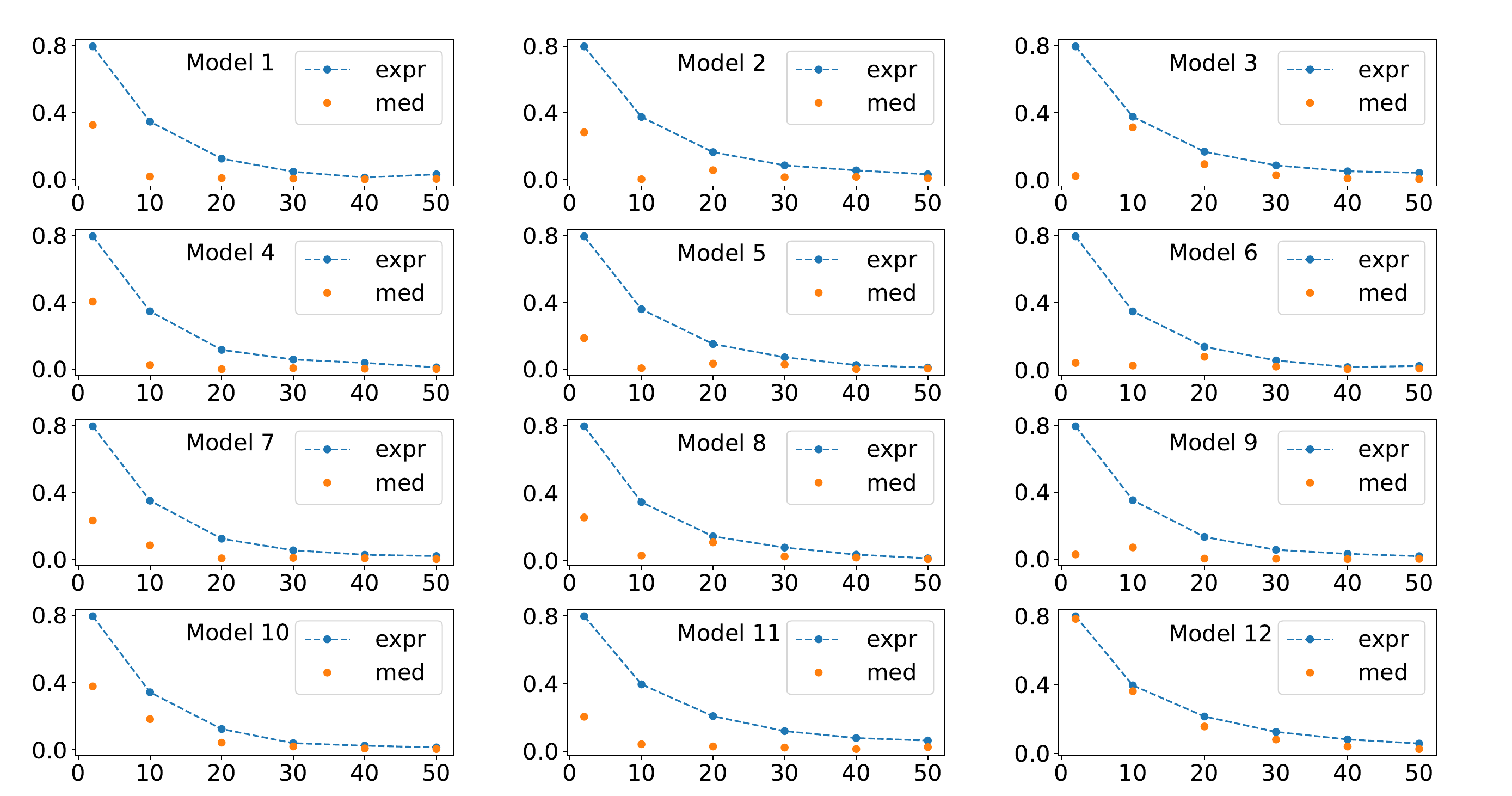}
    \caption{
     Behavour of the right hand side of Eq. (\ref{eq:eq_teo_1}), the quantum expressivity (expr), and of the average cost function term (med), the left hand side of Eq. (\ref{eq:eq_teo_1}),
     as the number of layers $L$ is increased. Five qubits were used for obtaining all these plots.}
    \label{fig:teo_1_5_qubit}
\end{figure*}
\begin{figure*}
    \centering
    \includegraphics[scale=0.40]{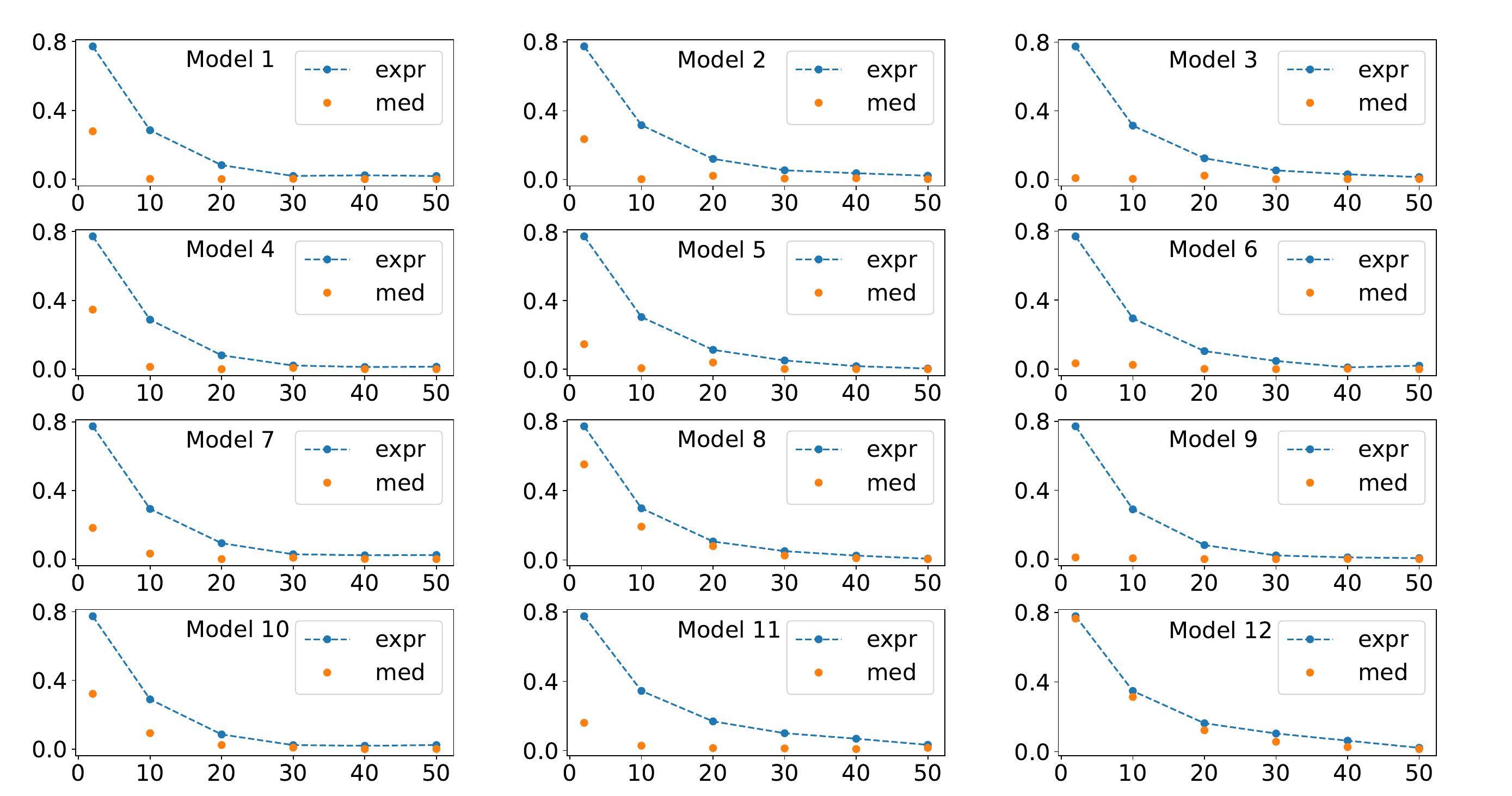}
    \caption{
    Behavour of the right hand side of Eq. (\ref{eq:eq_teo_1}), the quantum expressivity (expr), and of the average cost function term (med), the left hand side of Eq. (\ref{eq:eq_teo_1}),
    as the number of layers $L$ is increased. Six qubits were used for obtaining all these plots.}
    \label{fig:teo_1_6_qubit}
\end{figure*}

Initially, we numerically analyze Eq. \eqref{eq:eq_teo_1} of Theorem \ref{tr:concentracao}. For this, we performed an initial set of simulations, Figs. \ref{fig:teo_1_4_qubit}, \ref{fig:teo_1_5_qubit}, and \ref{fig:teo_1_6_qubit}, where we fixed the number of qubits and varied the number of layers $L$. For the results of Figs. \ref{fig:teo_1_4_qubit},
\ref{fig:teo_1_5_qubit}, and \ref{fig:teo_1_6_qubit}, we used four, five, and six qubits, respectively. Furthermore, for these simulations we consider the particular case $O = |0 \rangle \langle 0|$ and $\rho = |0 \rangle \langle 0|$.

Initially, we analytically calculate the value of $ \| A(\rho) \|_{2} $, where we get \cite{BR_expressibility}
\begin{equation}
    \| A(\rho) \|_{2} = \sqrt{ \mu(\rho) -1/d },\label{eq:expre_1}
\end{equation}
with 
\begin{equation}
    \mu(\rho) = \int_{\pmb{\theta}}\int_{\pmb{\phi}} | \langle \psi_{\pmb{\theta}} |  \psi_{\pmb{\phi}}  \rangle |^{2} d\pmb{\theta} d\pmb{\phi}.
\end{equation}
Or, from Ref. \cite{Sim_Expressibility}, we obtain
\begin{equation}
    \mu(\rho) = E[F] \text{     with     }  F =  | \langle \psi_{\pmb{\theta}} |  \psi_{\pmb{\phi}}  \rangle |^{2}.
\end{equation}

So, to calculate $\| A(\rho) \|_{2}$ we generated 5000 pairs of state vectors. Although we have generated a large number of state vectors, it is still a small sample of the entire Hilbert space.
So, the value we obtained for $\mu(\rho)$ is an approximation. As a consequence, in some simulations we obtained a complex value for $ \| A(\rho) \|_{2}$, Eq. \eqref{eq:expre_1}. Therefore, whenever this occurred, we restarted the simulation. Furthermore, we also used 5000 units to average the cost function.

In Figs. \ref{fig:teo_1_4_qubit}, \ref{fig:teo_1_5_qubit}, and \ref{fig:teo_1_6_qubit} is shown the behavour of the right hand side of Eq. (\ref{eq:eq_teo_1}), related to the expressivity, and of the average cost function term, the left hand side of Eq. (\ref{eq:eq_teo_1}). For producing these figures, four, five, and six qubits quantum circuits were used, respectively.

\begin{figure*}
    \centering
    \includegraphics[scale=0.40]{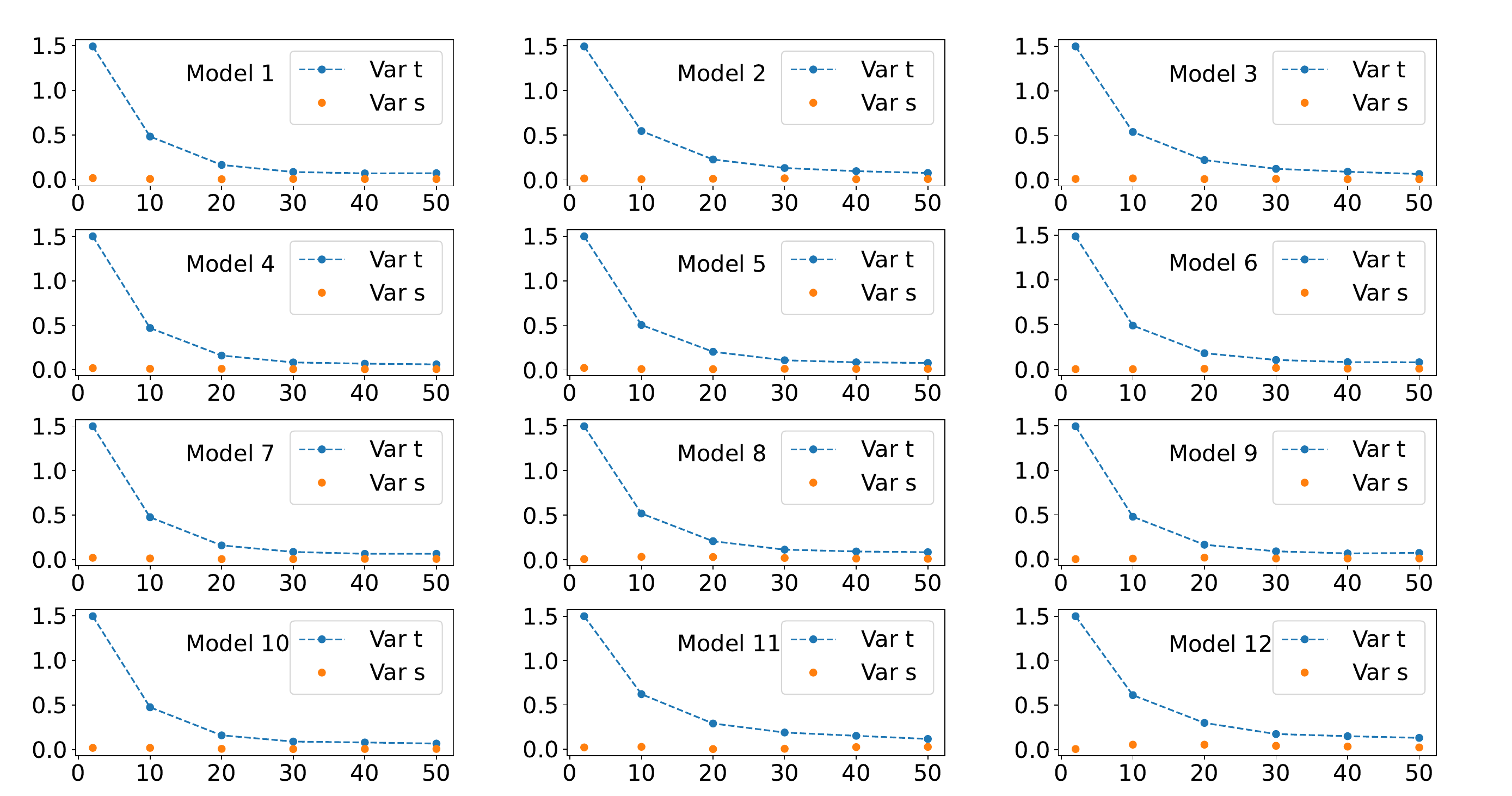}
    \caption{
    Behavour of the numerically calculated cost function variance, Var s, the left hand side of Eq. (\ref{eq:var_teorico}), and of the expressivity-related term, Var t, the right hand side of Eq. (\ref{eq:var_teorico}), as the number of layers $L$ is increased. Four qubits were used for obtaining all these plots.}
    \label{fig:var_4_qubit}
\end{figure*}

\begin{figure*}
    \centering
    \includegraphics[scale=0.40]{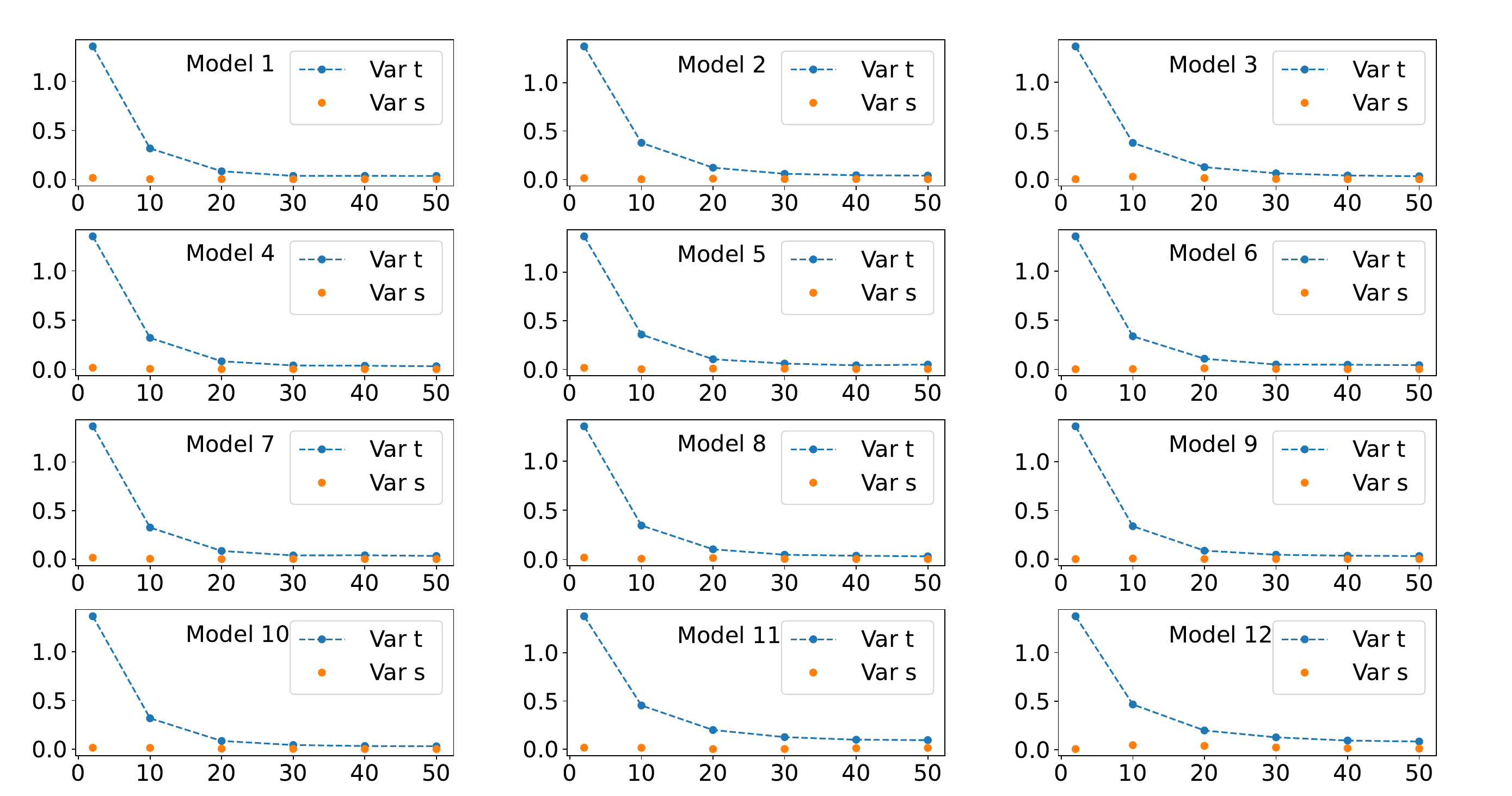}
    \caption{
    Behavour of the numerically calculated cost function variance, Var s, the left hand side of Eq. (\ref{eq:var_teorico}), and of the expressivity-related term, Var t, the right hand side of Eq. (\ref{eq:var_teorico}), as the number of layers $L$ is increased. Five qubits were used for obtaining all these plots.}
    \label{fig:var_5_qubit}
\end{figure*}

\begin{figure*}
    \centering
    \includegraphics[scale=0.40]{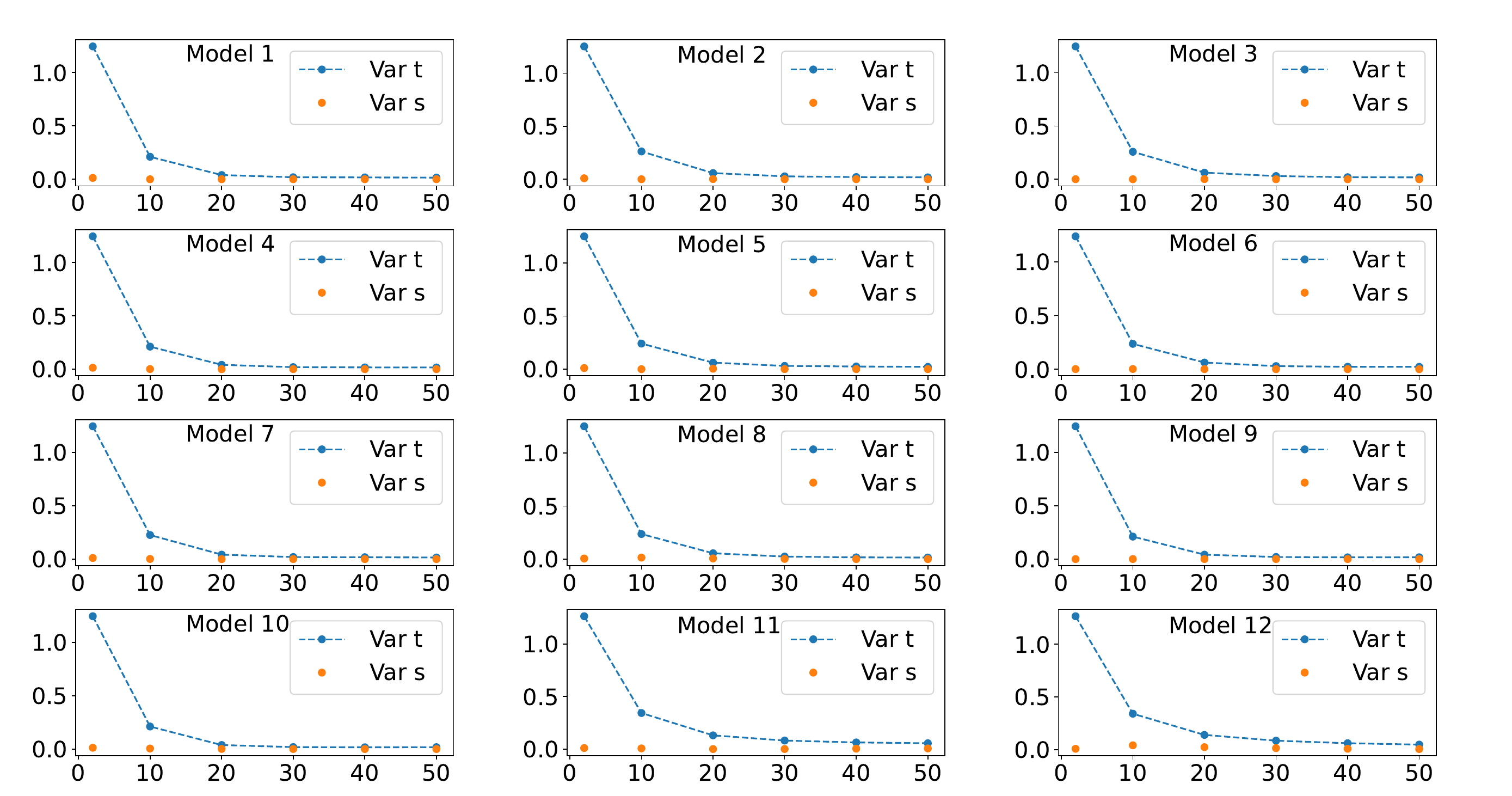}
    \caption{
    Behavour of the numerically calculated cost function variance, Var s, the left hand side of Eq. (\ref{eq:var_teorico}), and if the expressivity-related term, Var t, the right hand side of Eq. (\ref{eq:var_teorico}), as the number of layers $L$ is increased. Six qubits were used for obtaining all these plots.}
    \label{fig:var_6_qubit}
\end{figure*}

In Figs. \ref{fig:var_4_qubit}, \ref{fig:var_5_qubit}, and \ref{fig:var_6_qubit}, we show the behavior of the numerically calculated variance, Var s, the left hand side of Eq. (\ref{eq:var_teorico}), and of the theoretical value, Var t, the right hand side of Eq. \eqref{eq:var_teorico}, where we again used four, five, and six qubits, respectively. Also, we again used 5000 unitaries to compute the averages.

\clearpage 

\section{Conclusion}\label{sec:conclusion}

In this article, we analyzed how the expressiveness of the parametrization affects the cost function. As we observed, the concentration of the average value of the cost function has an upper limit that depends on the expressiveness of the parametrization, where, the more expressive this parametrization is, the more the average of the cost function will be concentrated around the fixed value $Tr[O]/d$ , Theorem \ref{tr:concentracao}. Furthermore, the probability for the cost function to deviate from its mean also depends on the expressiveness of the parametrization, Theorem \ref{tr:variancia}.

A possible implication of these results is related to the training of VQAs with highly expressive parametrizations. Once the more expressive the parametrization is, the more the average value of the cost function will be concentrated around $Tr[O]/d$, and the probability of the cost function deviating from this average also decreases, considering that, for the case where $ \| A(\rho)^{t} \|_{2} = 0$, the cost function will be stuck at the value $Tr[O]/d$. This result is in agreement with the one obtained in Ref. \cite{BR_expressibility}, where it was shown that the phenomenon of gradient disappearance is related with parametrization having high expressivity.

Another possible implication of our results is related to quantum machine learning models. In Ref. \cite{Hubregtsen_Thomas}, the authors mentioned that there is a correlation between expressiveness and accuracy, where the greater is the expressiveness, in general, the greater is the accuracy. To this end, the authors used Person's correlation coefficient to quantify this correlation. However, our results imply that, not only is the training of highly expressive parametrized quantum machine learning models difficult, as it will suffer more from the problem of gradient disappearance, as indicated in Ref. \cite{BR_expressibility}, but also the cost function itself will become stuck to a region close to the value $Tr[O]/d$.

\begin{acknowledgments}
This work was supported by the National Institute for the Science and Technology of Quantum Information (INCT-IQ), process 465469/2014-0, and by the National Council for Scientific and Technological Development (CNPq), processes 309862/2021-3 and 409673/2022-6.
\end{acknowledgments}

\appendix

\section{Proof of Theorem 1}
\label{ap:Appendix A}

For the proofs we present in this and in the next appendix, 
we shall need the following lemmas:
\begin{lemma}
\label{lemma_1}
Let $\{W_{y} \}_{y\in Y} \subset U(d)$ form a unitary t-design with $t \geq  1$, and let $A, B : H_{w} \rightarrow H_{w}$ be arbitrary linear
operators. Then
\begin{equation}
    \int d\mu(W)Tr[WAW^{\dagger}B] = \frac{Tr[A]Tr[B]}{d}. \label{eq:identity_1}
\end{equation}
\end{lemma}

\begin{lemma}
\label{lemma_2}
Let $\{W_{y} \}_{y\in Y} \subset U(d)$ form a unitary t-design with $t \geq  2$, and let $A, B, C, D : H_{w} \rightarrow H_{w}$ be arbitrary linear
operators. Then
\begin{equation}
    \begin{split}
    \int d\mu(W)Tr[WAW^{\dagger}B]Tr[WCW^{\dagger}D] = \\ = \frac{ Tr[A]Tr[B]Tr[C]Tr[D]+Tr[AC]Tr[BD] }{d^{2}-1} \\ - \frac{ Tr[AC]Tr[B]Tr[D]+Tr[A]Tr[C]Tr[BD] }{d(d^{2}-1)}, \label{eq:identity_2}
    \end{split}
\end{equation}
\end{lemma}
where $d=2^{n}$. For more details on the proofs of these lemmas, see Refs. \cite{Harr_measure_1,Harr_measure_2}.

We will also use the definition of expressivity given in Eq. (\ref{eq:expr_definition}). 
For proving Theorem \ref{tr:concentracao}, we start by writing the mean of the cost function as in Eq. (\ref{eq:media_cost_function}) 
Using $A(\rho):=A^{1}_{\mathbb{U}}(\rho)$ given by Eq. (\ref{eq:expr_definition}) with $t=1$ in Eq. \eqref{eq:media_cost_function}, we get
\begin{equation}
     E_{\mathbb{U}}[C] =  \int d\mu(V)Tr[OV\rho V^{\dagger}] - Tr[OA(\rho)].\label{eq:cost_function_apendice_A_1_1}
\end{equation}
Using the cyclicity of the trace function and Lemma \ref{lemma_1}, we obtain
\begin{equation}
     E_{\mathbb{U}}[C] =  \frac{Tr[O]}{d} - Tr[OA(\rho)],\label{eq:cost_function_apendice_A_1}
\end{equation}
where we used that $Tr[\rho] = 1$. From Eq. \eqref{eq:cost_function_apendice_A_1_1}, it follows that
\begin{equation}
    \begin{split}
        \bigg|E_{\mathbb{U}}[C] - \frac{Tr[O]}{d} \bigg|
        & =  |Tr[OA(\rho)]| \\
         & \leqslant  \| O \|_{2} \|A(\rho) \|_{2}.
        \label{eq:cost_function_apendice_A_1}
    \end{split} 
\end{equation}
Above we used the Cauchy-Schwarz inequality. With this, we have proved Theorem \ref{tr:concentracao}.

\section{Proof of Theorem 2}
\label{ap:Appendix B}

In this section, we will prove Theorem \ref{tr:variancia}. To do so, we start by writing the variance as
\begin{equation}
    Var_{\mathbb{U}}[C] = E_{\mathbb{U}}[C^{2}] - E_{\mathbb{U}}[C]^{2},
\end{equation}
where $E_{\mathbb{U}}[C]$ has already been obtained in Eq. \eqref{eq:cost_function_apendice_A_1_1}. Therefore, we get $E_{\mathbb{U}}[C^{2}]$, which will be given by
\begin{equation}
    \begin{split}
    E_{\mathbb{U}}[C^{2}] & = \int dU (Tr[OU\rho U^{\dagger}])^{2} \\ &  = \int dU Tr[O^{\otimes 2} U^{\otimes 2} \rho^{\otimes 2} (U^{\dagger})^{\otimes 2}].\label{eq:media_c_2}
    \end{split}
\end{equation}

Thus, using $A^{\otimes 2}(\rho^{\otimes 2})$ in Eq. \eqref{eq:media_c_2}, with $A^{\otimes 2}=A^{2}_{\mathbb{U}}$, we get
\begin{equation}
    \begin{split}
    E_{\mathbb{U}}[C^{2}] &= \int d\mu(V)Tr[O^{\otimes 2} V^{\otimes 2} \rho^{\otimes 2} (V^{\dagger})^{\otimes 2}] \\ &- Tr[O^{\otimes 2}A^{\otimes 2}(\rho^{\otimes 2})].\label{eq:media_c_2_1}
     \end{split}
\end{equation}

To solve the integral that appears in Eq. \eqref{eq:media_c_2_1}, we use Lemma \ref{lemma_2}. So
\begin{equation}
\begin{split}
& \int d\mu(V)Tr[O V \rho V^{\dagger} ] Tr[OV\rho V^{\dagger}] \\
& = \frac{Tr[O]^{2}+Tr[O^{2}]}{d^{2}-1} - \frac{Tr[O]^{2}+Tr[O^{2}]}{d(d^{2}-1)} \\
& = \frac{Tr[O]^{2}+Tr[O^{2}]}{d^{2}-1}\bigg( 1 - \frac{1}{d} \bigg),
\end{split}
\end{equation}
where we again used the cyclicity of the trace function. We also used $Tr[\rho] = 1$ and $Tr[\rho^{2}] = 1$.

Using this result in Eq. \eqref{eq:media_c_2_1}, we get
\begin{equation}
    E_{\mathbb{U}}[C^{2}] =  \frac{Tr[O]^{2}+Tr[O^{2}]}{d^{2}-1}\bigg( 1 - \frac{1}{d} \bigg) -  Tr[O^{\otimes 2}A^{\otimes 2}(\rho^{\otimes 2})].\label{eq:c_2}
\end{equation}

Using the results obtained in Eq. \eqref{eq:cost_function_apendice_A_1} and \eqref{eq:c_2}, we have that the variance will be given by
\begin{equation}
    \begin{split}
     Var_{\mathbb{U}}[C] & = \frac{Tr[O]^{2}+Tr[O^{2}]}{d^{2}-1}\bigg( 1 - \frac{1}{d} \bigg) -  Tr[O^{\otimes 2}A^{\otimes 2}(\rho^{\otimes 2})] \\
     & - \bigg[ \frac{Tr[O]}{d} - Tr[OA(\rho)] \bigg]^{2}
      \end{split}
\end{equation}
or
\begin{equation}
    \begin{split}
     Var_{\mathbb{U}}[C] & = \beta -  Tr[O^{\otimes 2}A^{\otimes 2}(\rho^{\otimes 2})] \\
     & + \alpha Tr[OA(\rho)]  - Tr[OA(\rho)]^{2} \label{eq:var_f_1}
      \end{split}
\end{equation}
with 
\begin{equation}
    \beta :=  \frac{Tr[O]^{2}+Tr[O^{2}]}{d^{2}-1}\bigg( 1 - \frac{1}{d} \bigg)
\end{equation}
and
\begin{equation}
    \alpha := \frac{2Tr[O]}{d}.
\end{equation}

Therefore, from Eq. \eqref{eq:var_f_1}, we get
\begin{equation}
    \begin{split}
     |Var_{\mathbb{U}}[C]| & \leqslant |\beta| +  |Tr[O^{\otimes 2}A^{\otimes 2}(\rho^{\otimes 2})]| \\
     & \hspace{0.4cm}+ |\alpha Tr[OA(\rho)]|  + |Tr[OA(\rho)]^{2}| \\
     & \leqslant |\beta| + \|O^{\otimes 2}\|_{2}\|A^{\otimes 2}(\rho^{\otimes 2})\|_{2} \\
     & \hspace{0.4cm} + |\alpha| \|O\|_{2}\|A(\rho)\|_{2} + \|O\|_{2}^{2}\|A(\rho)\|_{2}^{2} \label{eq:var_f_2},
      \end{split}
\end{equation}
where the first inequality was obtained from the triangular inequality for complex numbers and the second inequality is obtained from the Cauchy-Schwarz inequality. In addition, we use $|x^{2}| = |x|^{2} $ for $x\in\mathbb{R}$. With this, we complete the  proof of Theorem \ref{tr:variancia}.

\section{Appendix C}
\label{ap:Appendix C}

For all simulations that we presented in Sec. \ref{sec:result}, we used the parametrizations shown in the figures below.

\begin{figure*}[h]
    \centering
    \includegraphics[scale=0.5]{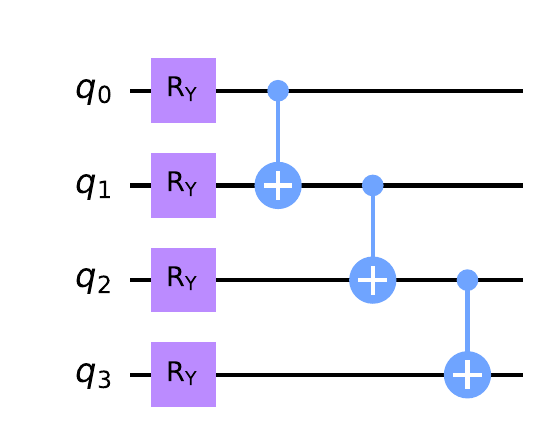}
    \caption{Model 1.}
    \label{fig:figura_model_1}
\end{figure*}

\begin{figure*}[h]
    \centering
    \includegraphics[scale=0.5]{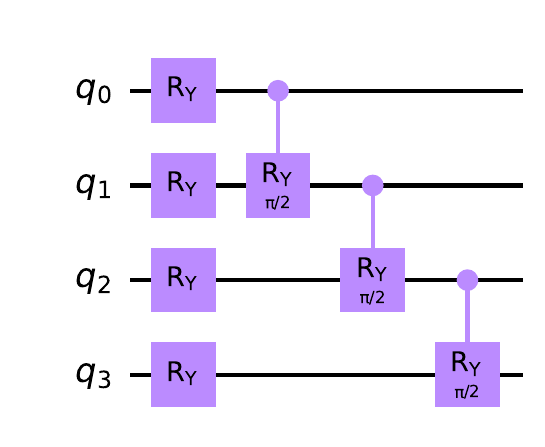}
    \caption{Model 2.}
    \label{fig:figura_model_2}
\end{figure*}

\begin{figure*}[h]
    \centering
    \includegraphics[scale=0.5]{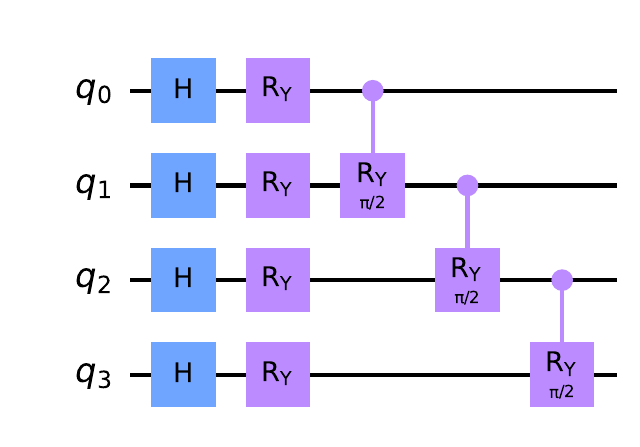}
    \caption{Model 3.}
    \label{fig:figura_model_3}
\end{figure*}

\begin{figure*}[h]
    \centering
    \includegraphics[scale=0.5]{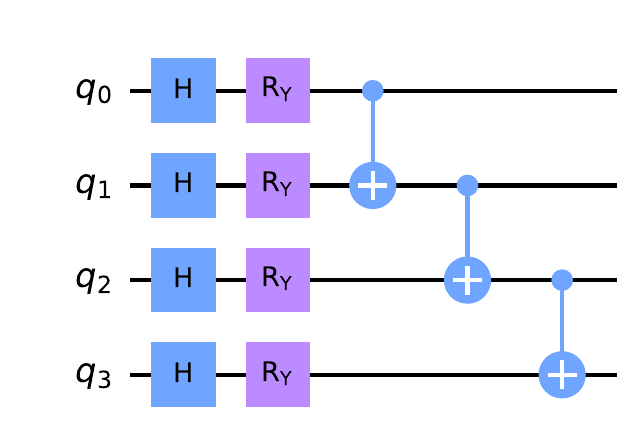}
    \caption{Model 4.}
    \label{fig:figura_model_4}
\end{figure*}

\begin{figure*}[h]
    \centering
    \includegraphics[scale=0.5]{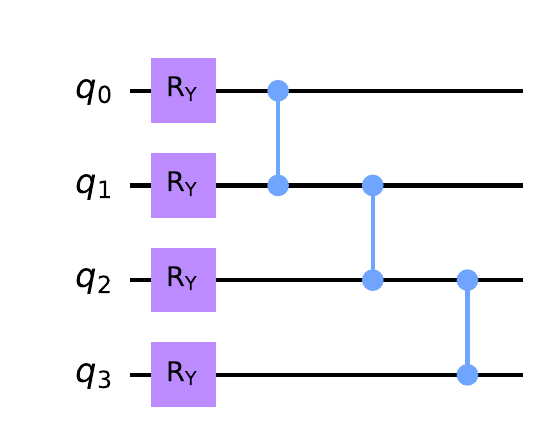}
    \caption{Model 5.}
    \label{fig:figura_model_5}
\end{figure*}

\begin{figure*}[h]
    \centering
    \includegraphics[scale=0.5]{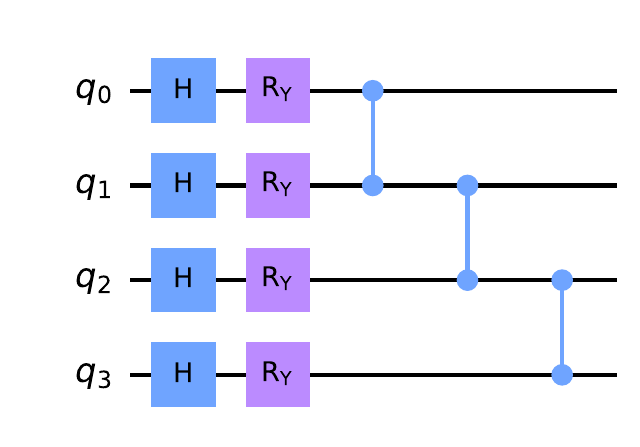}
    \caption{Model 6.}
    \label{fig:figura_model_6}
\end{figure*}

\begin{figure*}[h]
    \centering
    \includegraphics[scale=0.5]{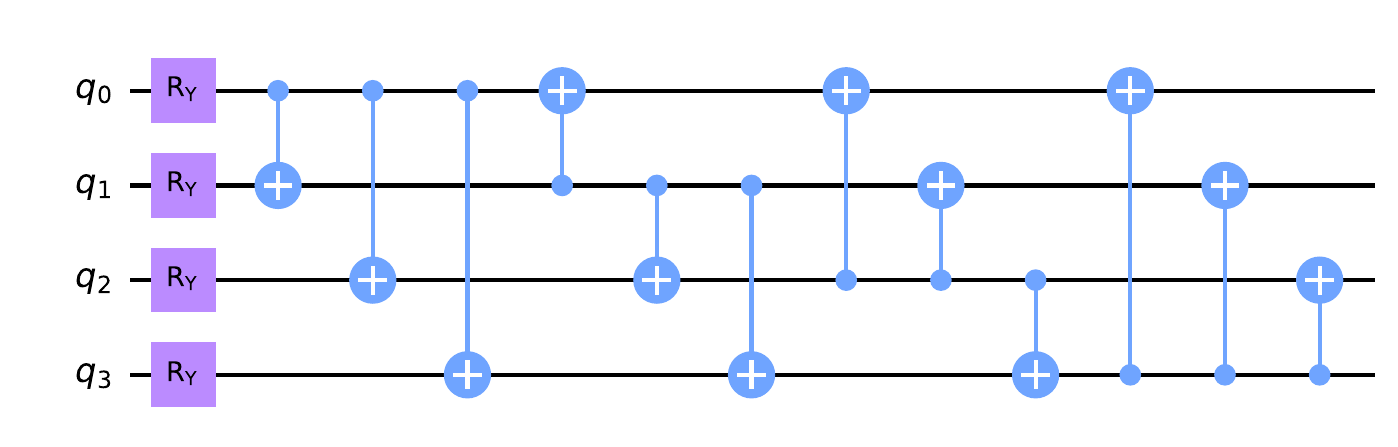}
    \caption{Model 7.}
    \label{fig:figura_model_7}
\end{figure*}

\begin{figure*}[h]
    \centering
    \includegraphics[scale=0.5]{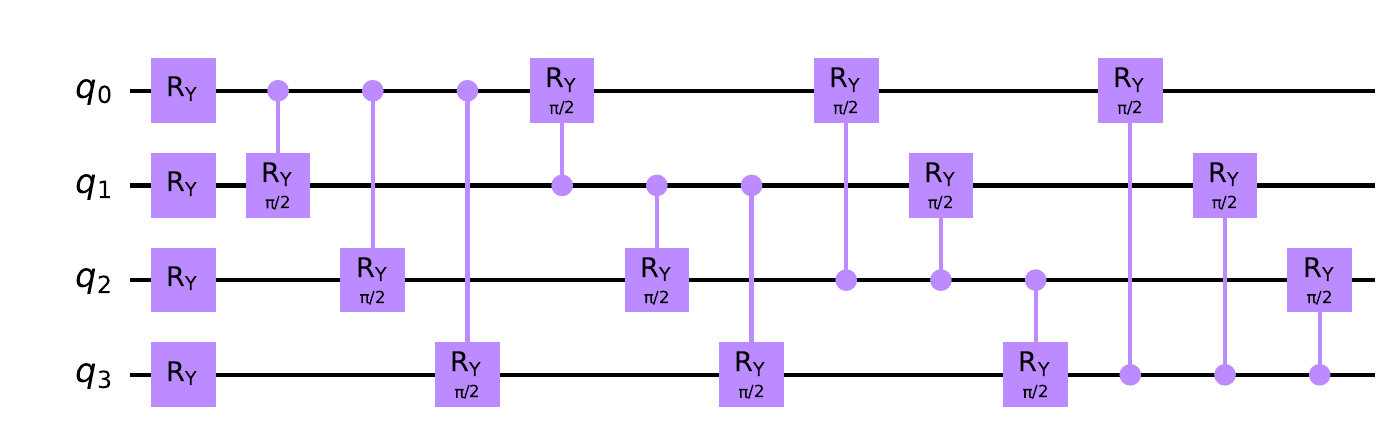}
    \caption{Model 8.}
    \label{fig:figura_model_8}
\end{figure*}

\begin{figure*}[h]
    \centering
    \includegraphics[scale=0.5]{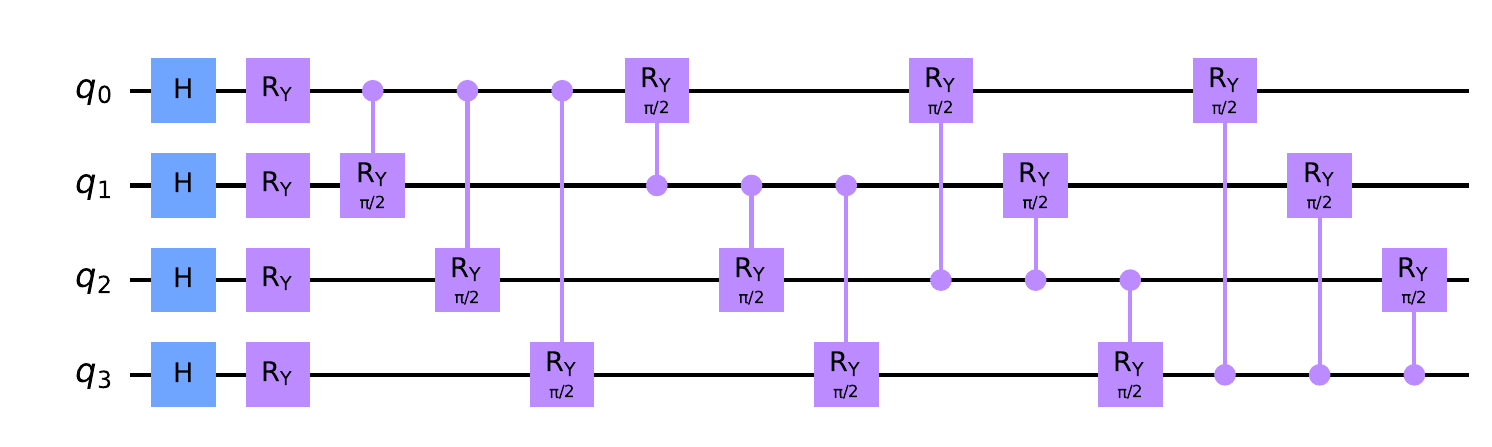}
    \caption{Model 9.}
    \label{fig:figura_model_9}
\end{figure*}

\begin{figure*}[h]
    \centering
    \includegraphics[scale=0.5]{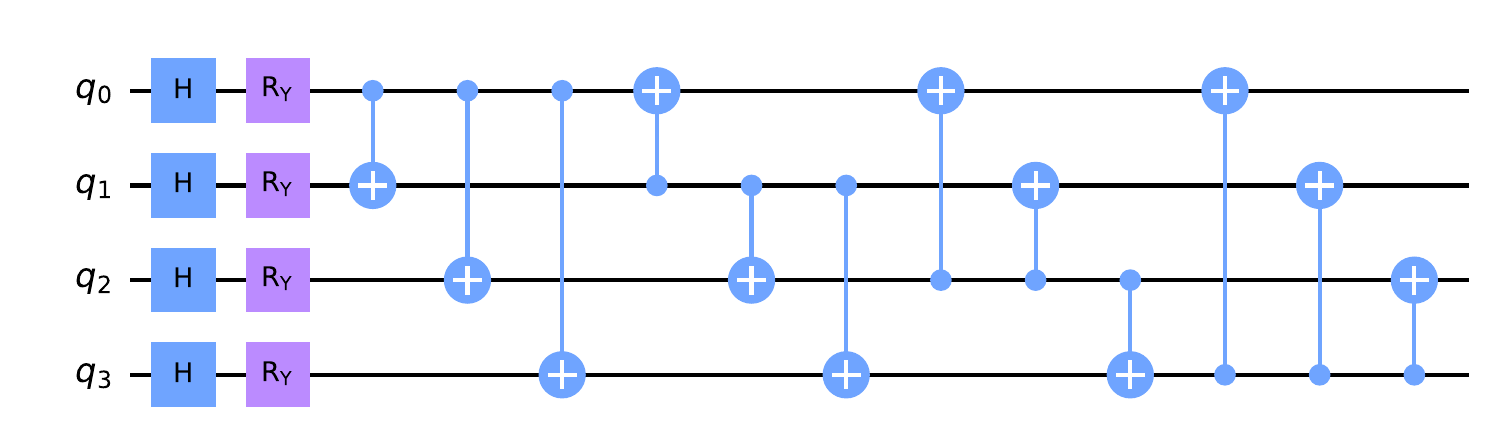}
    \caption{Model 10.}
    \label{fig:figura_model_10}
\end{figure*}

\begin{figure*}[h]
    \centering
    \includegraphics[scale=0.5]{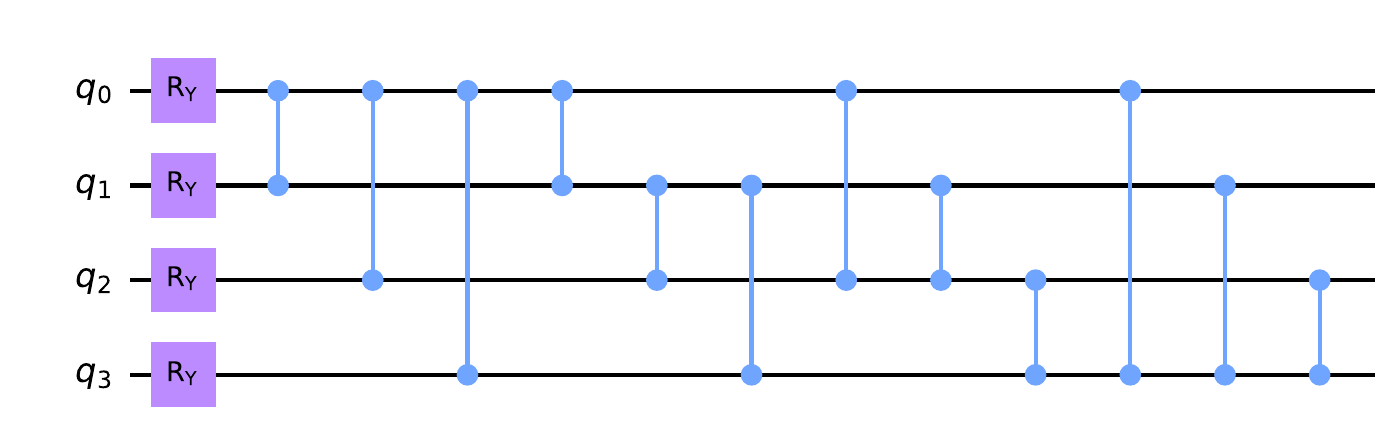}
    \caption{Model 11.}
    \label{fig:figura_model_11}
\end{figure*}

\begin{figure*}[h]
    \centering
    \includegraphics[scale=0.5]{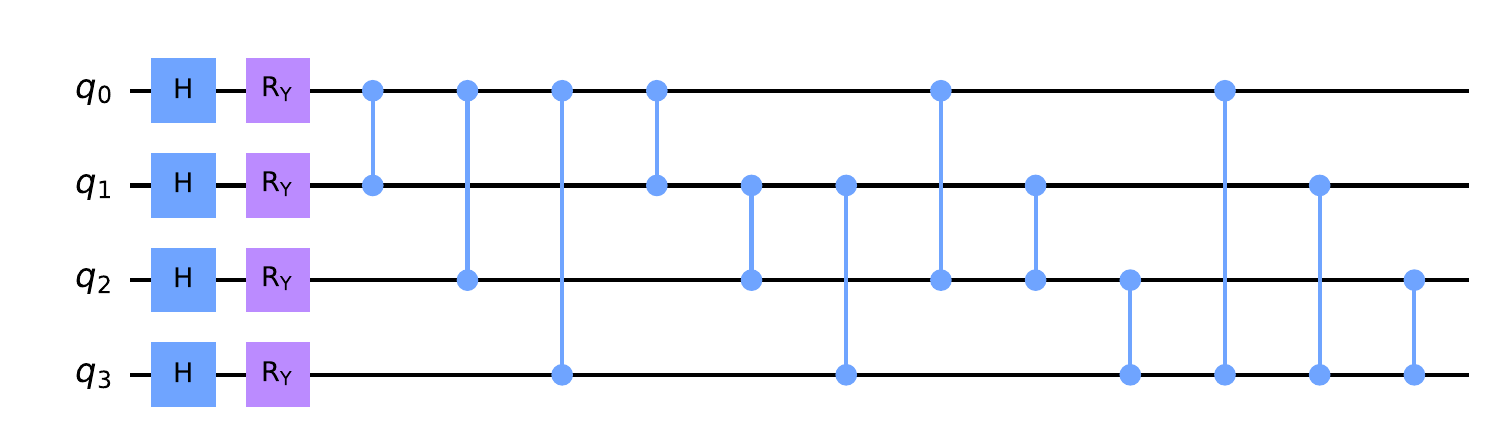}
    \caption{Model 12.}
    \label{fig:figura_model_12}
\end{figure*}

\end{document}